\input{aipcheck}

\documentclass[
    ,final            
  ]
  {aipproc}

\layoutstyle{6x9}

\usepackage{amssymb,amsmath,array}

\newcommand{\FLpcac}{ F_L^{\rm PCAC} }
\newcommand{\FLac}{ F_L^{\rm A} }

\newcommand{\FLvc}{ F_L^{\rm V} }

\begin{document}

\title{Modeling Lepton-Nucleon Inelastic Scattering from 
High to Low Momentum Transfer}

\classification{13.15.+g  13.60.-r}     
\keywords      {Neutrino cross-sections, high twists, structure functions, parton distribution functions}

\author{S.~Alekhin}{
  address={Institute for High Energy Physics, 142281 Protvino, Moscow region, Russia}
}

\author{S.A.~Kulagin}{
  address={Institute for Nuclear Research, 117312 Moscow, Russia}
}

\author{R.~Petti}{
  address={Depertment of Physics and Astronomy, University of South Carolina, Columbia SC 29208, USA}
}

\begin{abstract}

We present a model for inclusive charged lepton-nucleon
and (anti)neutrino-nucleon cross sections at momentum transfer squared,  
$Q^2$, $\sim1~{\rm GeV}^2$. We quantify the impact of 
existing low-Q charged-lepton deep-inelastic scattering (DIS) data 
on effects due to high-twist operators and on the extraction of parton distribution functions (PDFs).  
No evidence is found for twist-6 contributions to structure functions (SF), 
and for a twist-4 term in the logitudinal SF at $x\gtrsim0.1$. 
We find that DIS data are consistent with the NNLO QCD approximation 
with the target mass and phenomenological high twist corrections.   
For $Q^2<1~{\rm GeV}^2$, we extend extrapolation of the operator product 
expansion, preserving the low-$Q$ current-conservation theorems. 
The procedure yields a good description of data down to $Q^2\sim 0.5~{\rm GeV}^2$.  
An updated set of PDFs with reduced uncertainty and applicable down to 
small momentum transfers in the lepton-nucleon scattering is obtained.

\end{abstract}

\maketitle


\section{Introduction}

At high momentum transfer $Q$ the lepton-nucleon cross-sections are 
well described in terms of parton distributions (PDFs), whose $Q^2$ 
evolution is well-understood in perturbative Quantum Chromo-Dynamics (QCD).    
The universality of the partonic description allows to obtain predictions for a 
variety of probes ($e, \mu, \nu, \bar{\nu}$) and targets, which have been extensively 
verified by experiments. However, by lowering progressively $Q$ non-perturbative 
phenomena become more and more important for a precise modeling of cross-sections. 
  
The existing DIS data at small momentum transfer $Q$ in principle can  
shed light on the interplay between perturbative and non-perturbative 
phenomena and clarify the limits of applicability of the parton model.  
Furthermore they can put valuable constraints on the parton distribution functions (PDFs)
and on the strong coupling constant $\alpha_{\rm s}$, due to their excellent statistical precision. 
However, significant high-order QCD corrections are required to study such kinematical region. 
The recent progress in the NNLO QCD calculations~\cite{NNLO} allows to use the DIS data down 
to $Q\sim 1~{\rm GeV}$ in global QCD fits by keeping the perturbative stability under control. 

In the formalism of the Operator Product Expansion (OPE) unpolarized structure 
functions can be expressed in terms of powers of $1/Q^2$ (power corrections):  
\begin{equation}
F_{2,T,3}(x,Q^2) = F_{2,T,3}^{\tau = 2}(x,Q^2)
+ {H_{2,T,3}^{\tau = 4}(x) \over Q^2} 
+ {H_{2,T,3}^{\tau = 6}(x) \over Q^4} + .....   
\label{eqn:ht}
\end{equation}
The first term ($\tau=2$), expressed in terms of PDFs, represents the 
Leading Twist (HT) describing the scattering off a free quark and 
is responsible for the scaling of SF via perturbative QCD 
$\alpha_s(Q^2)$ corrections. The Higher Twist (HT) terms ($\tau = 4,6$) 
reflect instead the strength of multi-parton correlations (qq and qg). 
Since such corrections spoil factorization one 
has to consider their impact on the PDFs extracted in the analysis of low-$Q$ data.  
Due to their non-perturbative origin, current models can only provide a 
qualitative description for such contributions, which must then be determined 
phenomenologically from data.   

Existing information about high twist terms in lepton-nucleon structure functions is 
scarce and somewhat controversial. Early analyses~\cite{Miramontes:1989ni,Rslac} 
suggested a significant HT contribution to the longitudinal SF $F_L$. The subsequent 
studies with both charged leptons~\cite{Virchaux92,BodekYang00,Alekhin02} and 
neutrinos~\cite{Kataev99} raised the question 
of a possible dependence on the order of QCD calculation used for the leding twist.  
The common wisdom is generally that HTs only affect the region of $Q^2 \sim 1 \div 3$ GeV$^2$  
and can be neglected in the extraction of the leading twist.  
  
In this communication we report our results on using the DIS data 
down to $Q=1~{\rm GeV}$ in the global QCD fit of PDFs with power 
corrections included in the analysis.

\section{Procedure}

The analyzed data set consist of the world charged-leptons DIS
cross section data for the proton and deuteron targets
by the SLAC, BCDMS, NMC, FNAL-E-665, H1, and ZEUS experiments
supplemented by the fixed-target Drell-Yan data,
the latter constrain the sea quark distribution, which is
poorly determined by the DIS data alone. Basically the same combination
of data was used in the earlier fit of Ref.\cite{Alekhin:2006zm} with
the cut $Q^2>2.5~{\rm GeV}^2$ imposed on the DIS data.
In the present fit alongside with the softer cut imposed on the
SLAC and NMC data,
$Q>1~{\rm GeV}$, we also add the DIS data by FNAL-E-665
experiment~\cite{Adams:1996gu} since they give additional constraint
on the PDFs at small $x$ provided not too stringent cut on $Q$ is
applied. The cut on invariant mass of the hadron system
$W>1.8~{\rm GeV}$ is imposed on the DIS data to avoid the 
resonance region. The total number of data points
(NDP) used in the fit is 3076, in the range of $x=0.0001\div0.9$.
The analysis~\footnote{Details of the theoretical ansatz can be found
in Ref.\protect{\cite{Alekhin:2006zm}}.}  
is performed in the NNLO QCD approximation
with the target mass corrections~\cite{TMC} taken into account and the
dynamical twist-4 (twist-6) terms parameterized in the additive form 
as model independent spline functions $H(x)$.   
Deuteron data are corrected for nuclear effects following 
the model~\cite{Kulagin:2004ie}. 

In addition, the recent neutrino and antineutrino cross-section 
data from the CHORUS experiment~\cite{chorus-xsec} are added to the 
global fit for $Q>1.0$ GeV and $x\geq0.045$, mainly to constrain the corresponding HT terms.

\section{Determination of High Twist Terms}

As a first step of the analysis we checked possible twist-6 contributions  
to the DIS structure functions by keeping the terms $H^{\tau=6}(x)$ in Eq.(\ref{eqn:ht}).
Due to the $W$ cut, twist-6 terms are insensitive to the large-$x$ data 
and therefore we set them to zero at $x\geq 0.5$.  
Values of the HT terms at $x=0$ were also set to zero in view of
the fact no clear evidence of saturation effects was found in the small $x$ HERA data.

\begin{figure}
  \includegraphics[height=.35\textheight]{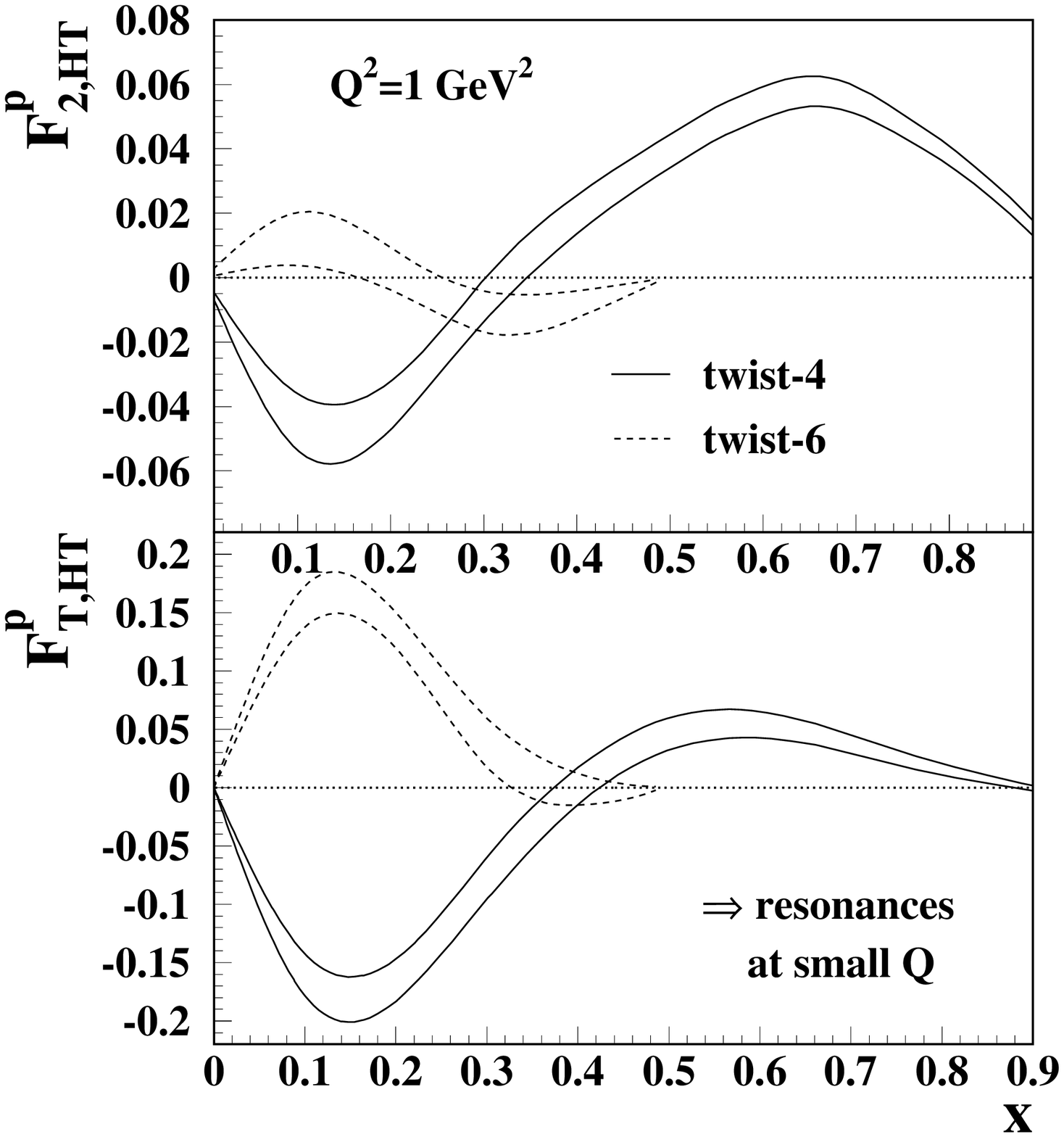}\includegraphics[height=.35\textheight]{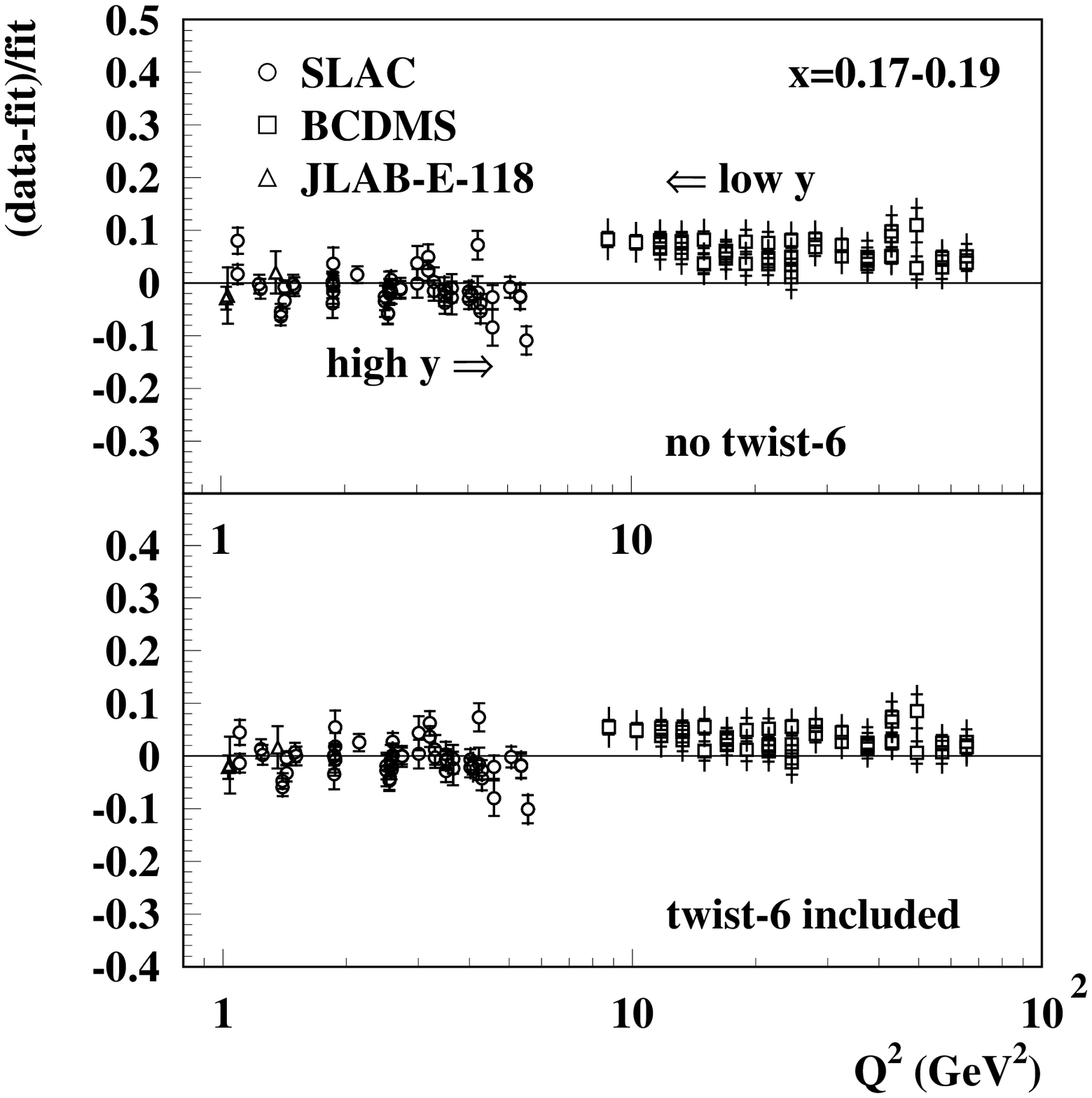}
  \caption{Left figure: the $1\sigma$ error bands for the twist-4 (solid lines) and twist-6 (dashes) 
terms in the proton $F_2$ (upper panel) and $F_{\rm T}$ (lower panel). 
The arrow indicates the region of $x$ with the limited potential for the determination of 
twist-6 terms due to the cut on $W$. 
Right figure: pulls corresponding to the fits with and without twist-6 terms.  
The arrows in the upper panel indicate the high-$y$ and low-$y$ regions for the SLAC and BCDMS data. 
The data points for the JLAB-E-118 experiment at $Q\sim1~{\rm GeV}$ were not used in the fit.
}\label{fig:twist6}  
\end{figure}

\begin{figure}
  \hspace*{-0.60cm}\includegraphics[height=.35\textheight]{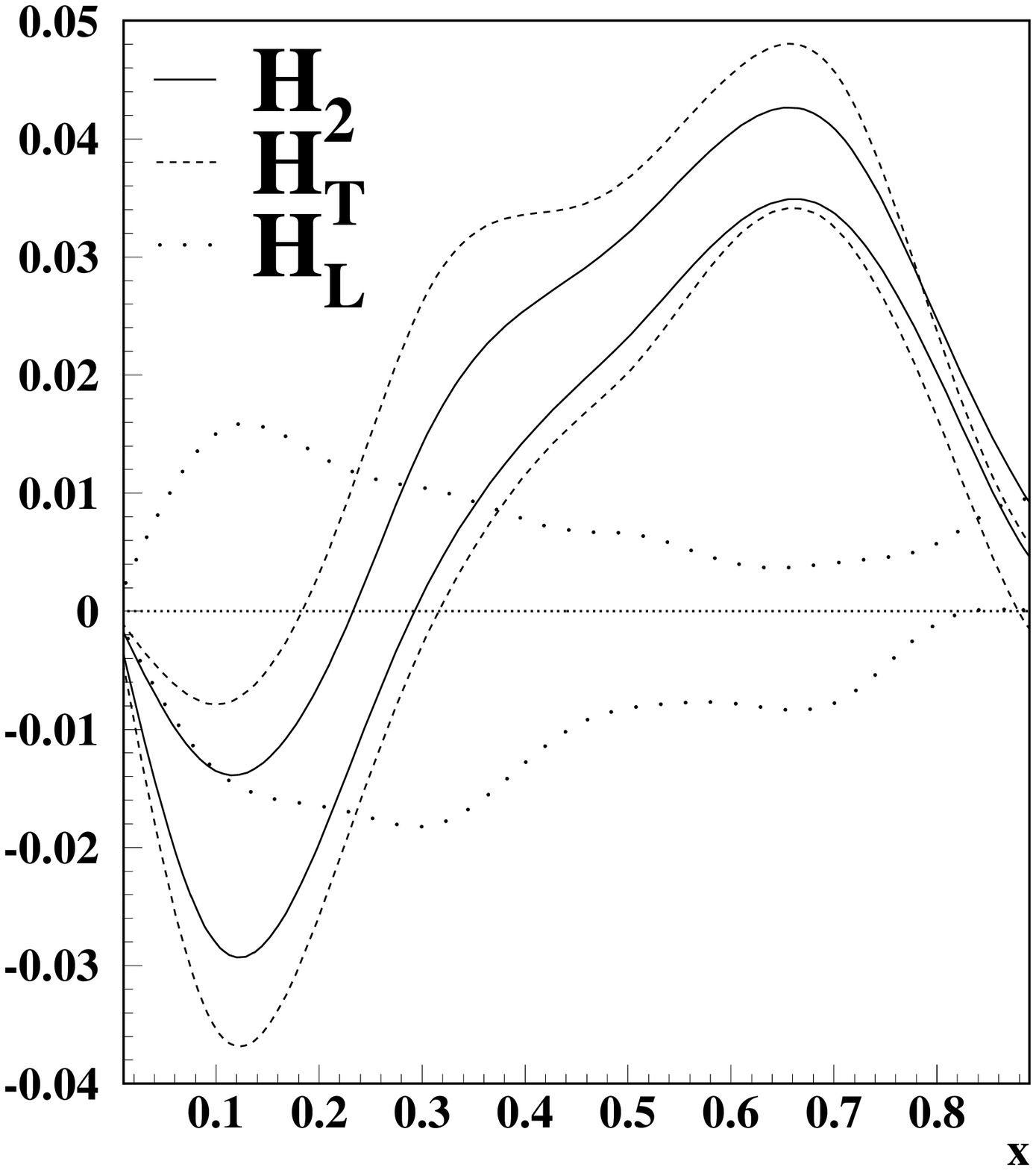}\includegraphics[height=.35\textheight]{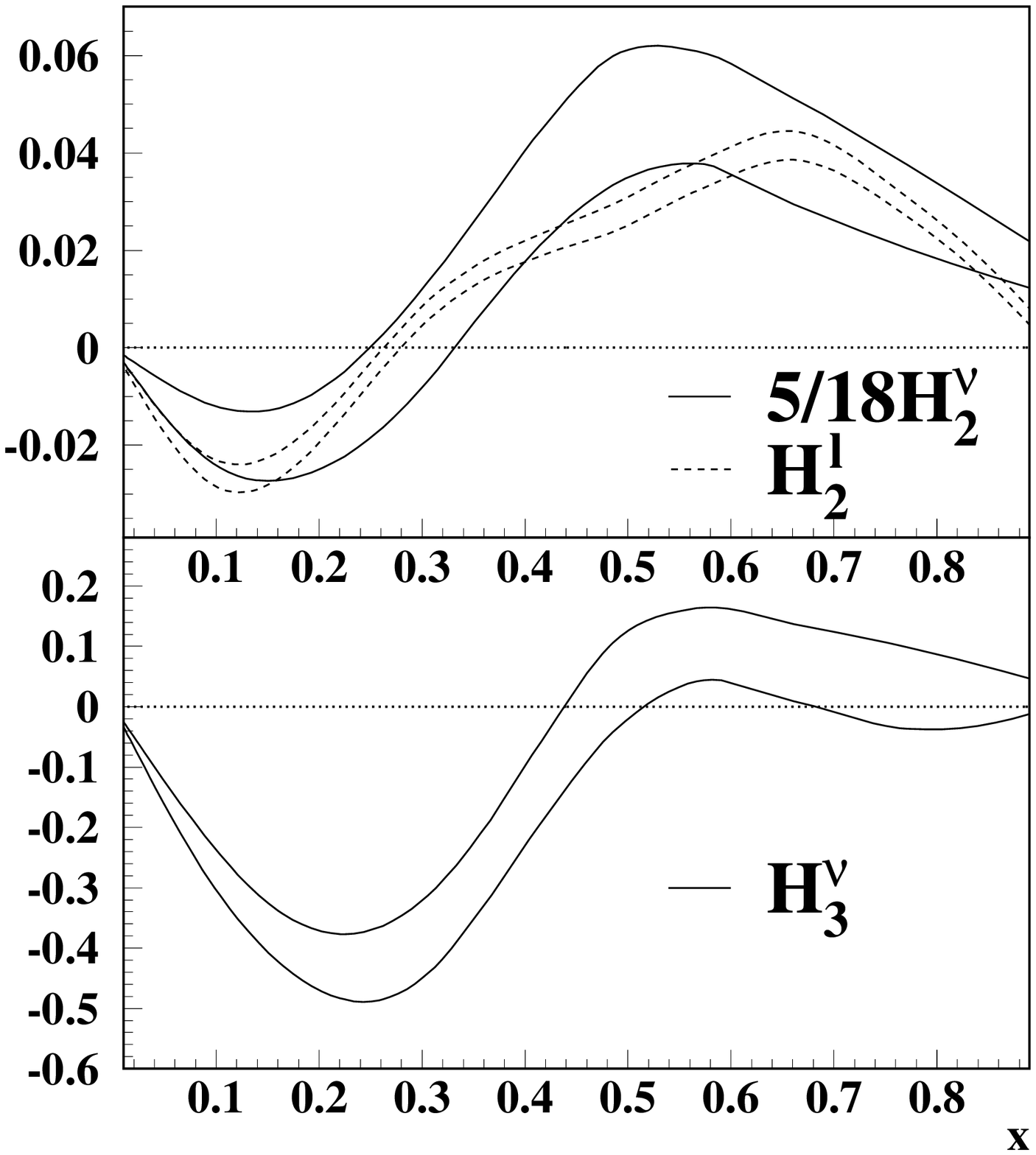}
  \caption{Left figure: the $1\sigma$ error bands for the high-twist terms in the isospin-symmetric combinations of  
different structure functions (solid lines: $F_2$, dashes: $F_{\rm T}$, dots: $F_{\rm L}$) for charged leptons.
Right figure: corresponding $1\sigma$ bands for neutrino scattering off an isoscalar target (upper panel: $F_2$, 
lower panel: $xF_3$). The predictions for $F_2$ from charged leptons rescaled by the corresponding leading twist terms 
are also shown for comparison.   
}\label{fig:finalHT}     
\end{figure}

Figure~\ref{fig:twist6} (left) shows the HT terms in $F_{\rm 2,T}$ obtained in such a variant of our fit.   
Surprisingly, we observe a large positive contribution from the twist-6 term to $F_{\rm T}$ at $x\sim 0.15$.
At the same time this contribution is compensated by a corresponding negative twist-4 term.
Since the twist-4 and twist-6 coefficients are similar in magnitude but  
opposite in sign, the overall sum of HT terms demonstrate a weak dependence on $Q$. 
This observation leads us to the conclusion that in fact the twist-6 term in $F_{\rm T}$
absorbs some non-power-like effects. 

When we remove twist-6 terms from our fits the magnitude of the pulls appears to be maximal 
not at the lowest $Q$ values, but at $Q^2\sim7~{\rm GeV}^2$, exactly in the region of overlap 
between SLAC and BCDMS data (see Fig.\ref{fig:twist6} right).
Indeed, these two data sets show a certain discrepancy, which is directly translated   
into a fake twist-6 contribution, if such term is fitted as well.
The values of inelasticity $y$ are large for SLAC data at the highest $Q$   
and small for the BCDMS data at the lowest $Q$. For this reason, the discrepancy 
affects mainly $F_{\rm T}$, which is more sensitive to $y$ than $F_2$.
On the other hand the SLAC/BCDMS inconsistency at $Q^2\sim7~{\rm GeV}^2$
is not crucial for the determination of twist-4 terms.
If we rescale the uncertainties of data in this region    
to bring the pulls at the level of $1\sigma$, we observe a negligible  
increase in the corresponding HT errors.    
Evidently, HT terms are driven by data at the lowest $Q$ available, so that   
in principle the SLAC/BCDMS data around $Q^2\sim7~{\rm GeV}^2$
can even be dropped without any loss of statistical power. Preliminary data from the 
experiment JLAB-E-118~\cite{E118} agree with the SLAC data at low $Q$, thus favouring the 
reliability of the latter at $Q\sim1~{\rm GeV}$, regardless of potential problems 
in the region of overlap with BCDMS.  

Coming to the conclusion that the twist-6 terms observed are
just an artefact due to certain inconsistencies in the data,  
we drop them from the final version of the fit.  
This results in a value of $\chi^2/{\rm NDP}$ of 3815/3076=1.25,
which is higher than unity in consideration of the data discrepancies discussed 
above. Nonetheless the problematic data points are spread out more or
less randomly over kinematics and they do not bias the results of the fit.  
A rescaling of the uncertainties in the data points with the largest pulls, such that 
the overall $\chi^2/{\rm NDP}$ becomes unity, leads to a modest increase in the errors 
of PDFs and HTs within 20\%. Figure~\ref{fig:finalHT} (left) shows the twist-4 
terms obtained in the final version of our fit. The HT contribution to $F_T$ is remarkably 
similar to the one in $F_2$, despite the two terms were fitted independently.
As a result the HT term in $F_L$ defined as $H_{\rm L}=H_2-H_{\rm T}$
is well comparable to zero within the uncertainties. In the final version of the fit 
we then impose the constraint $H_2=H_T$ for the isospin-symmetric combinations of 
structure functions.   

\begin{figure}
  \includegraphics[height=.35\textheight]{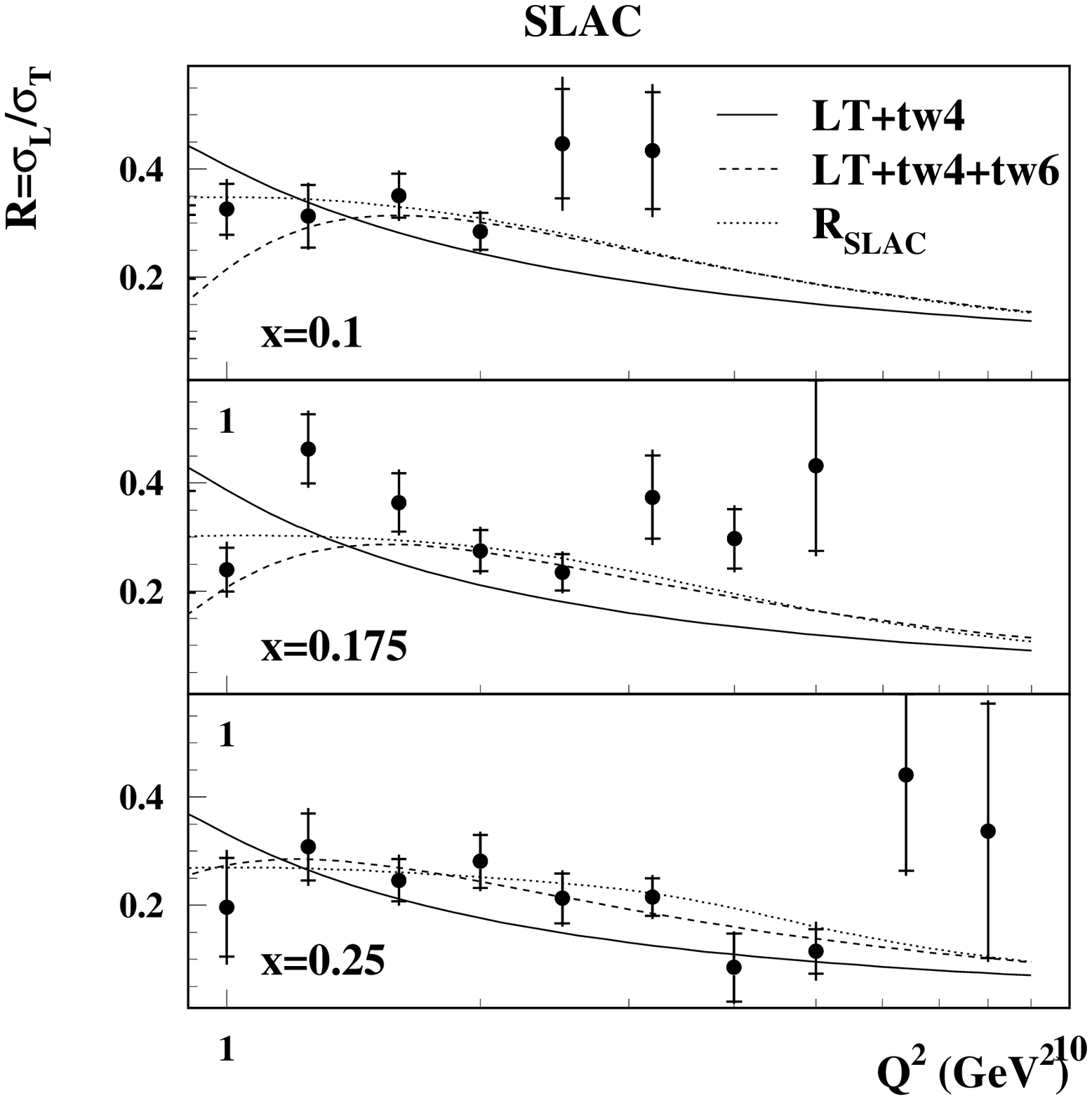}\includegraphics[height=.35\textheight]{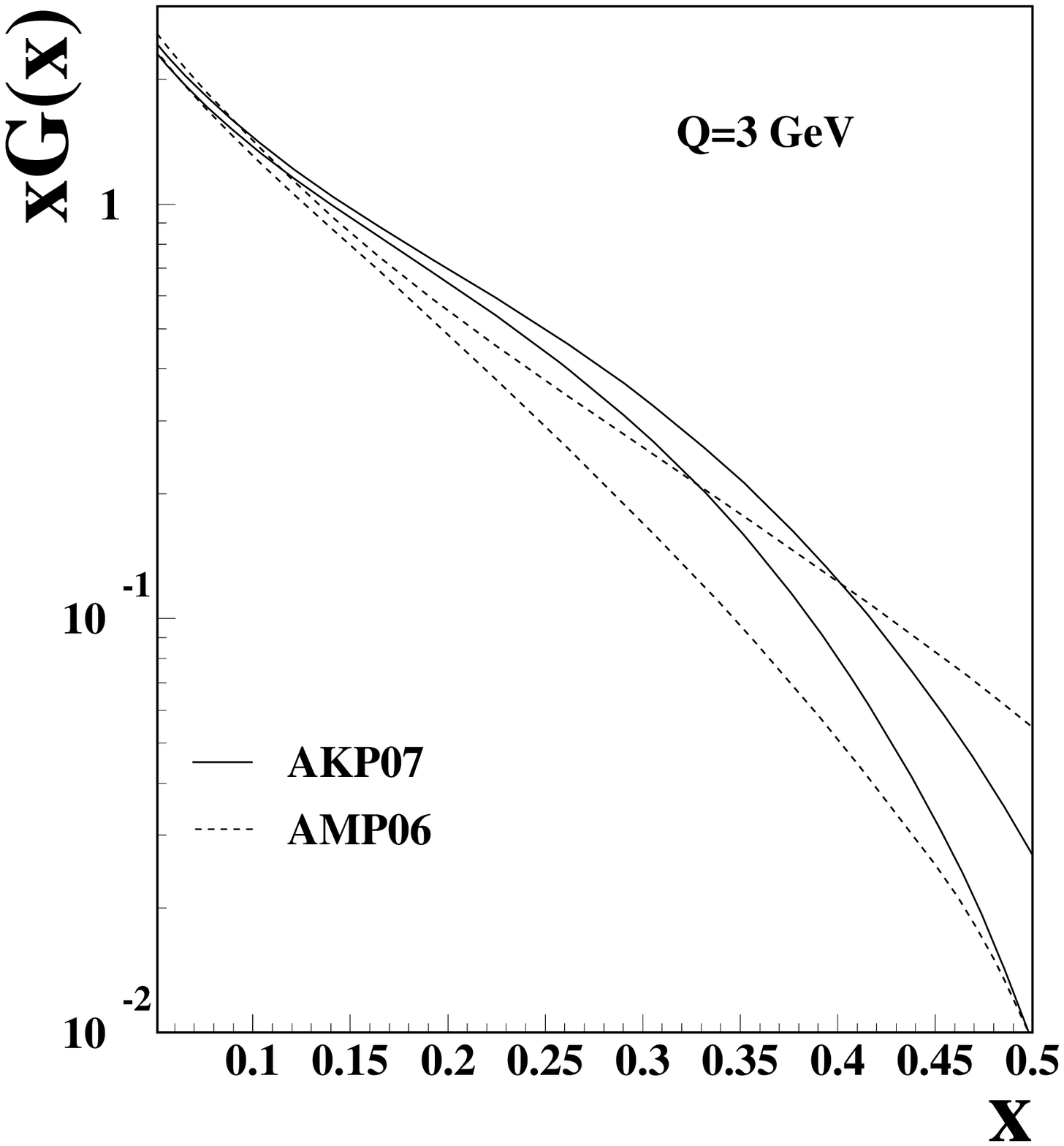}
  \caption{Left figure: comparison of the calculations for $R=\sigma_L/\sigma_T$ including high 
twist contributions with the empirical parameterization of SLAC data $R_{\rm SLAC}$~\cite{Rslac}. 
The data points represent SLAC data.
Right figure: the $1\sigma$ error bands for the gluon distribution obtained in our fit (solid lines) compared to 
one of Ref.~\cite{Alekhin:2006zm}(dashes).
}\label{fig:lowQeffects}
\end{figure}

Our results indicate the HT contribution to the structure function $R=\sigma_{\rm L}/\sigma_{\rm T}$ 
is also small in the whole range of $x$ considered. This is in contrast to the 
conclusion of Ref.\cite{Miramontes:1989ni}
about a large HT contribution to $R$. We explain such difference by the fact that the low-$Q$ part of the SLAC
data was not considered in Ref.\cite{Miramontes:1989ni}. The extrapolation of those results to the lower
$Q$ value must be in disagreement with data. In Fig.~\ref{fig:lowQeffects} (left) we compare our 
predictions for $R$ with the empirical parameterization $R_{\rm SLAC}$ from SLAC data~\cite{Rslac}.   
The latter is consistent with our calculation based upon the fit with fake twist-6 terms.   
This indicates the $R_{\rm SLAC}$ parameterization is the result of the same inconsistency between SLAC and 
BCDMS data we discussed above.  

It is interesting to determine the HT contributions to (anti)neutrino structure 
functions independently from the ones extracted from charged lepton data.   
Due to the structure of the weak Charged Current (CC) some similarities with charged leptons 
could be expected for $F_2$ and $F_T$. 
Since the target nucleon is mostly isoscalar for neutrino data we  
impose the constraint $H_2=H_T$, consistently with charged leptons~\footnote{This relation 
does not hold in general since the presence of an axial-vector current introduces significant 
HT contributions to $F_L$ in the limit of small $x$ values and vanishing $Q$ (PCAC). This 
will be discussed in the following. However, we are here focused on a kinematic region in which 
such effects are marginal ($x\geq0.045$ and $Q>1.0$ GeV).}.    
Figure~\ref{fig:finalHT} (right) summarizes our observations. The ratio $H_2^{\tau 4}/F_2^{\rm LT}$ 
is remarkably similar for both (anti)neutrinos and charged leptons over the entire $x$ range.  
In addition, the use of neutrino data allows us to extract simultaneously 
the HT contribution to $xF_3$. Our results indicate overall $H_3$ 
provides a negative contribution to the Gross--Llewellyn-Smith integral, which is 
consistent with the predictions of Ref.~\cite{BraunKolesnichenko87}.   

The total contribution of the HT terms into the DIS cross section turns out 
to be small compared to the leading-twist (LT) part. For a realistic DIS kinematics 
the ratio of the HT and LT terms is $\lesssim10\%$, which justifies
the use of the twist expansion in our analysis.

Finally, our results demonstrate high twist contributions to 
unpolarized structure functions do not vanish in the NNLO approximation.   
Indeed we find no strong dependence upon the order of QCD calculation used 
in the leading twist.   
   
\section{Impact on Leading Twist}

The use of additional data at low $Q$ values allows to achieve a better 
separation between leading and higher twists by exploiting their 
different $Q^2$ dependence. The correlation coefficients are indeed 
substantially reduced in the whole $x$ range by extending the lower 
cutoff from 2.5 GeV$^2$ to 1.0 GeV$^2$ and do not exceed 0.3. 
This in turn results in reduced theoretical uncertainties.     

Our value of $\alpha_{\rm s} (M_Z) = 0.1128 \pm 0.0011$ is consistent 
with the one obtained in Ref.\cite{Alekhin:2006zm} with the cut
$Q^2>2.5~{\rm GeV}^2$ and is in good agreement with the
result from the non-singlet DIS data analysis~\cite{Blumlein:2006be}.
The PDFs obtained in the fit with the low-$Q$ DIS data included are also 
close to the ones of Ref.\cite{Alekhin:2006zm}.
This manifests a good separation of the LT and HT terms
in the fit and a stability of the perturbative calculation. 
The most significant change is observed in the gluon distribution at
$x\sim0.3$, which is enhanced in the low-$Q$ fit (see Fig.\ref{fig:lowQeffects} right).
This is due to the significant twist-4 term appearing in $R$ in the fit
with $Q^2>2.5~{\rm GeV}^2$, similarly to the analysis of Ref.\cite{Miramontes:1989ni}.   
The LT terms in $R$ is then correspondingly suppressed. Since the LT in $R$
is proportional to the value of the gluon distribution, 
the latter is also suppressed as a result.

The uncertainties on PDFs are improved as compared to the fit of
Ref.\cite{Alekhin:2006zm}. In particular, the $d$-quark distribution
is now determined within few per cent at $x\sim0.2$, 
which is comparable to the precision of the $u$-quark distribution.
This improvement has important phenomenological consequences for the 
extraction of the weak mixing angle from neutrino data. 
For instance the analysis by the NuTeV collaboration
\cite{Zeller:2001hh}, based on the Paschos--Wolfenstein relation, 
requires a good knowledge of the valence distributions in order   
to guarantee an accurate correction for the target 
non-isoscalarity~\cite{Kulagin:2003wz}.
This correction is proportional to the ratio $x_1/x_0$,
where $x_1$ and $x_0$ are the integrals over $x$ of the iso-vector
and iso-scalar combinations of the valence quarks, respectively.
From our fit we obtain $x_1/x_0=0.424(6)$, which provides an 
uncertainty on the non-isoscalarity of the target comparable 
to the experimental uncertainties from NuTeV data.

\section{Extrapolation to $Q<1$ GeV}  

In order to describe the structure functions in the region of low momentum transfer   
we apply a smooth interpolation between the high $Q^2$ regime, which is described 
in QCD in terms of LT and HT contributions as discussed above, and the $Q^2\to0$
predictions derived from current conservation arguments.
We use $Q^2_m = 1$ GeV$^2$ as the matching point between high- and
low-$Q^2$ regimes.

\begin{figure}
  \includegraphics[height=.35\textheight]{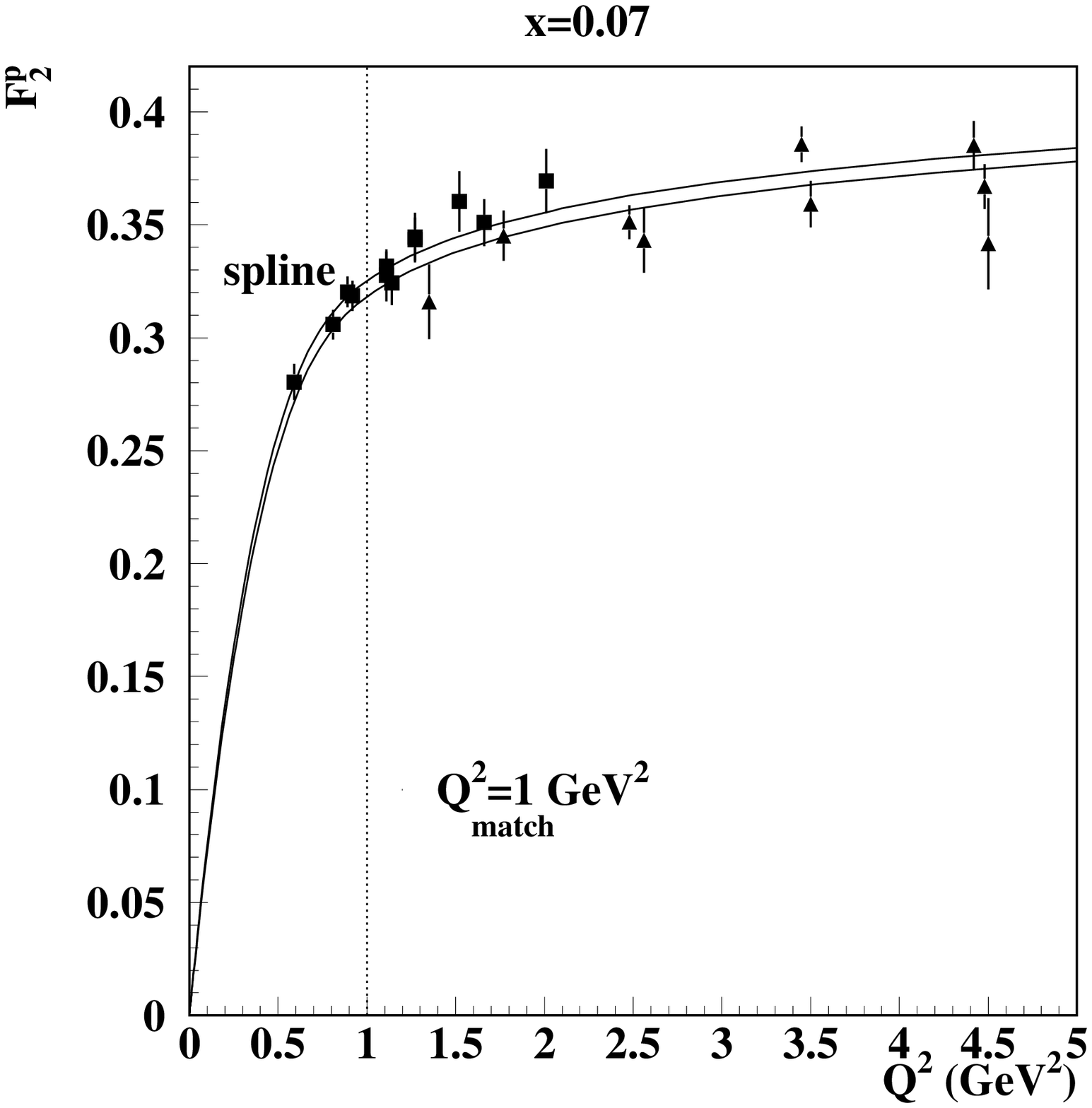}\includegraphics[height=.35\textheight]{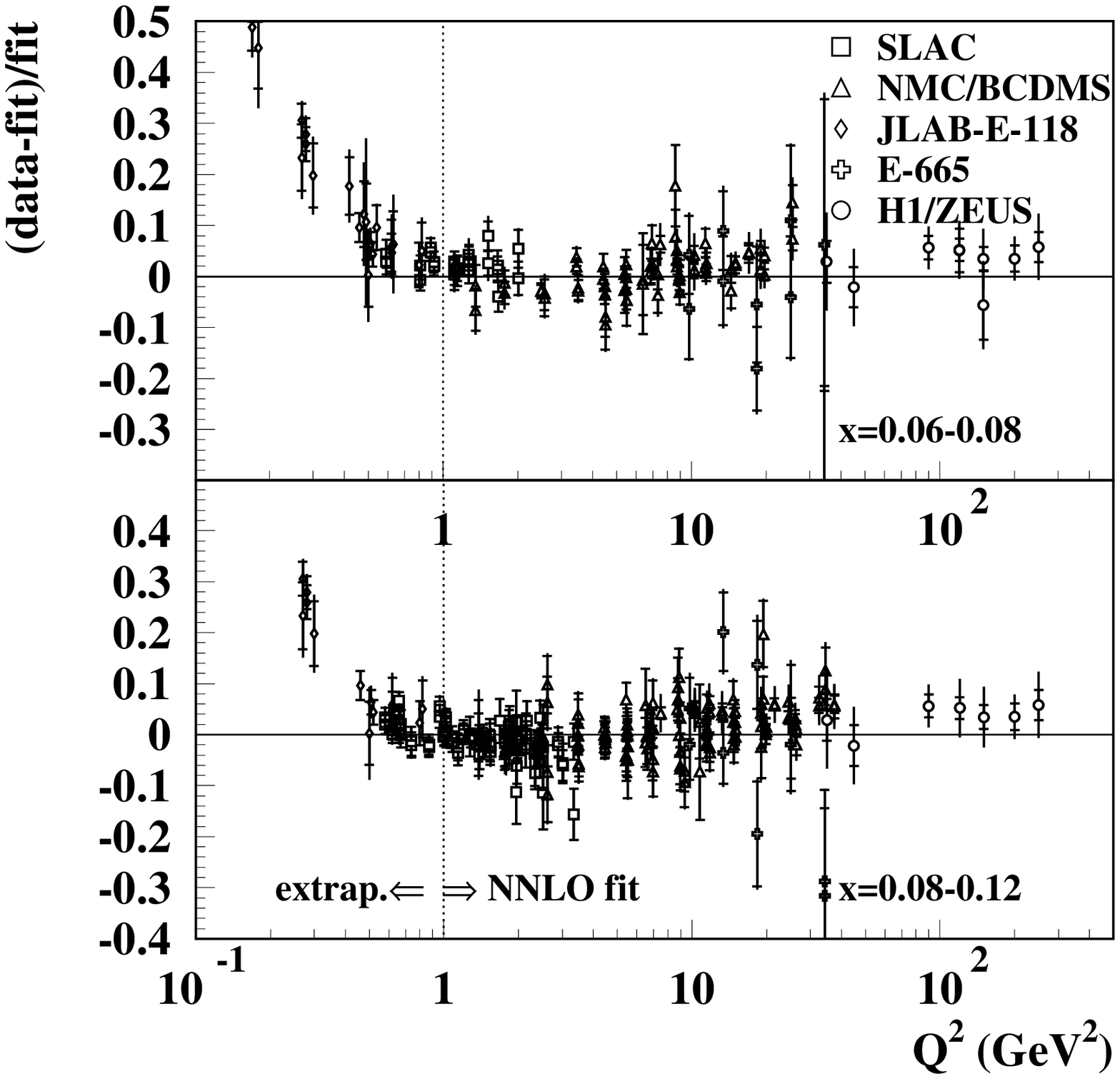}
  \caption{Left figure: interpolation of structure functions in the region $0<Q^2<1~GeV^2$.
The example given in the plot refers to $F_2$ for charged lepton scattering on
protons at x=0.07 (see text for details). 
Right figure: pulls with respect to the calculations from our fit for charged leptons as a function of $Q^2$. 
The region $Q^2<1$ GeV$^2$ and data from JLab E-118 were not included in the fit. 
}\label{fig:lowQextrap}  
\end{figure}

\begin{figure}
  \includegraphics[height=.35\textheight]{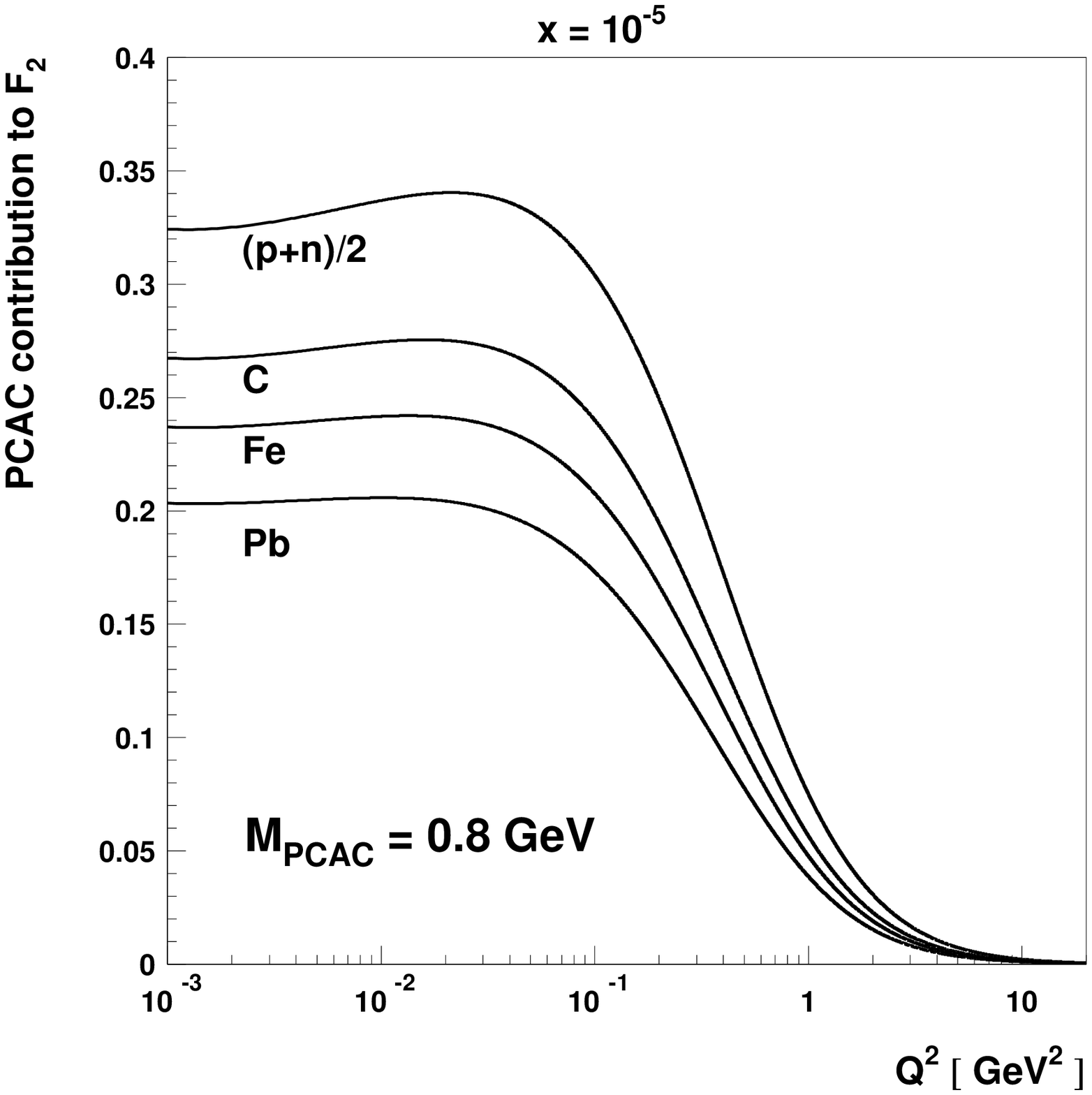}\includegraphics[height=.35\textheight]{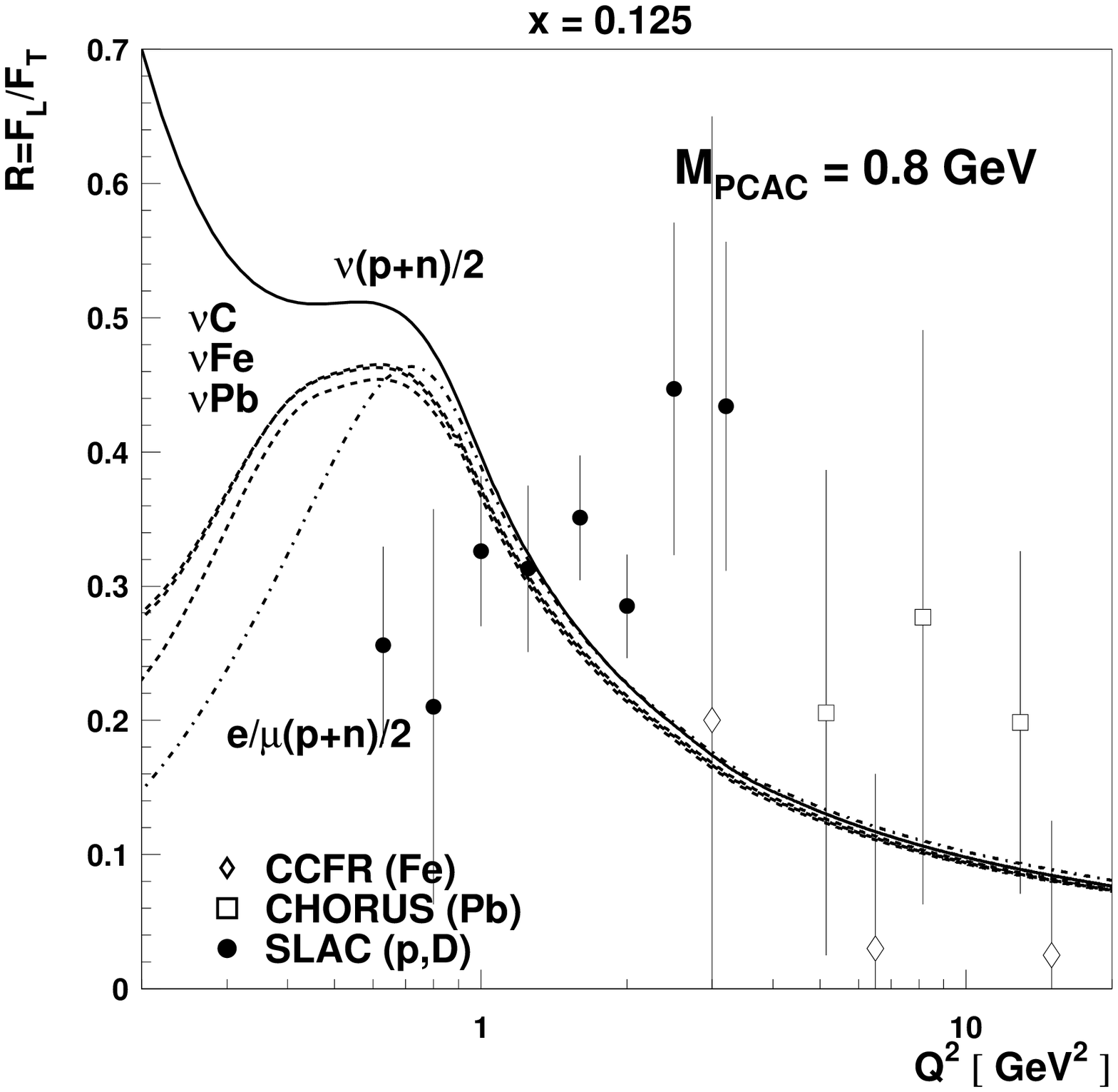}
  \caption{Left figure: the PCAC contribution to the neutrino structure function $F_2$ calculated for 
$x=10^{-5}$ as a function of $Q^2$ for a few different targets (labels on the curves). 
A value $M_{\textsc{pcac}}=0.8$ GeV is assumed.   
Right figure: comparison of the ratio $R=F_L/F_T$ calculated for the isoscalar
nucleon for the charged lepton (dashed-dotted line) and neutrino (solid
line) cases at $x=0.125$. Also shown are the results for different nuclear targets 
(${}^{12}$C, ${}^{56}$Fe and ${}^{207}$Pb from top to bottom). 
A value $M_{\textsc{pcac}}=0.8$~GeV is assumed. Determinations from
SLAC~\cite{Rslac}, CCFR~\cite{Yang:2001xc} and CHORUS~\cite{chorus-xsec}
are given for comparison.
}\label{fig:pcac}    
\end{figure}

The conservation of the vector current (CVC) suggests that for the electromagnetic 
interaction in the limit of $Q^2 \to 0$ the structure function $F_T$ vanishes as $Q^2$   
while $F_L$ vanishes faster than $F_T$ in such a way that $F_L/F_T \sim Q^2$.
Although these asymptotic conditions define the value of structure functions at $Q^2=0$,  
they do not specify at which scale such behaviour sets in. On the other hand, 
at $Q^2_m$ we know precisely the value of SFs from the twist expansion.    
In the region $0<Q^2<Q^2_m$ we interpolate by using cubic splines calculated for fixed $x$ values.
The corresponding coefficients are fully determined by the condition
both functions and derivatives should match with the twist expansion
at $Q^2_m$. Figure~\ref{fig:lowQextrap} (right) illustrates the interpolation procedure for
$F_2$ on protons in charged lepton scattering.   

We obtain a good description of charged lepton data down to $Q^2 \sim 0.5$ GeV$^2$, as can be seen 
from Fig.~\ref{fig:lowQextrap} (left) showing pulls as a function of $Q^2$ at 
a fixed $x$ value. Below that some deviations are visible in the recent 
JLab E-118 data. This may indicate the transition from dynamics to the behaviour 
dictated by gauge invariance is slower than in our simple extrapolation of 
the twist expansion.   

In the low-$Q$ region (anti)neutrino cross sections are
dominated by the longitudinal structure function $F_L$ and the latter is
driven by the axial-current interactions. Similarly to the charged lepton case, 
the structure function $F_T$ vanishes as $Q^2$ at low $Q^2$. This behaviour holds
for both the vector and the axial-vector contributions. However,
in the longitudinal channel the low-$Q$ behavior of the vector and
axial-vector parts are different. The vector component $\FLvc$ still vanishes 
as $Q^4$ at low $Q^2$ values.  

In contrast to the vector current, the axial-vector current is not
conserved. For low momentum transfer the divergence of the
axial-vector current is proportional to the pion field
(Partially Conserved Axial Current or PCAC): $\partial A^\pm = f_\pi m_\pi^2 \varphi^\pm$   
%
%
where $m_\pi$ is the pion mass, $f_\pi=0.93 m_\pi$ the pion decay
constant and $\varphi^\pm$ the pion field in the corresponding charge
state. We introduce explicitely a PCAC contribution to $\FLac$:
\begin{equation}
\label{FL:ac}
\FLac = \gamma^{3} \FLpcac f_{\rm PCAC}(Q^2)
+ \widetilde{F}_L^{\rm A}
\end{equation}
where $\gamma=(1+4x^2M^2/Q^2)^{1/2}$, $\FLpcac=f_\pi^2\sigma_\pi/\pi$ and
$\sigma_\pi=\sigma_\pi(s,Q^2)$
is the total cross section for the scattering of a virtual pion with
the center-of-mass energy squared $s$.
The last term $\widetilde{F}_L^{\rm A}$ is similar to $\FLvc$ and vanishes as $Q^4$.
Since the PCAC contribution is expected to vanish at high $Q^2$ we introduce a
form factor $f_{\rm PCAC}(Q^2)=(1+Q^2/M_{\rm PCAC}^2)^{-2}$~\cite{Kulagin:2007ie}, where the
dipole form is motivated by meson dominance arguments~\footnote{Formally the PCAC 
contribution can be considered as a high twist term.}. It is important to note
the pion pole does not directly contribute to structure functions and hence
the mass scale controlling the PCAC mechanism, $M_{\rm PCAC}$, cannot be
the pion mass itself, but is rather related to higher mass states like
$a_1$, $\rho \pi$ etc. This scale is not well known from theory and must be 
determined with data themselves. A value $M_{\rm PCAC}=0.8$ GeV seems to 
provide the best agreement with existing low-$Q$ data.    

Since $F_2=(F_L+F_T)/\gamma$ the presence of the PCAC terms implies the 
structure function $F_2$ at low $Q^2$ is dominated by the
nonvanishing $\FLpcac$ term. Figure~\ref{fig:pcac} (left) illustrates the magnitude of such contribution 
to $F_2$ for an isoscalar nucleon and for a few nuclear targets~\cite{Kulagin:2007ie}.  
The values of $F_2$ for heavy nuclear targets are systematically smaller because of 
nuclear shadowing effect for the pion cross section. Our predictions for the 
asymptotic value are consistent with the determination $F_2(Q^2=0)=0.210\pm0.002$ 
by the CCFR experiment on an iron target~\cite{ccfr-pcac}.    

It is instructive to compare the low-$Q^2$ behaviour of
$R=F_L/F_T$ for charged-lepton and neutrino scattering. In both cases
$F_T\propto Q^2$ as $Q^2\to0$. However, if $F_L\propto Q^4$ for the
electromagnetic current, for the weak current $F_L\to \FLpcac$ and thus
$F_L$ does not vanish in the low-$Q^2$ limit.
Then the behavior of $R$ at $Q^2\ll 1$ GeV$^2$ is substantially different for
charged-lepton and neutrino scattering. In Figure~\ref{fig:pcac} (right) 
we illustrate this effect by calculating $R$ as a function of $Q^2$ at  
a fixed $x$ for an isoscalar nucleon and a number of nuclei. 
We observe the nuclear correction partially reduces the differences 
between charged leptons and neutrinos and smoothes out the divergence 
of $R$ in the latter case.

\begin{figure}
  \hspace*{-0.90cm}\includegraphics[height=.40\textheight]{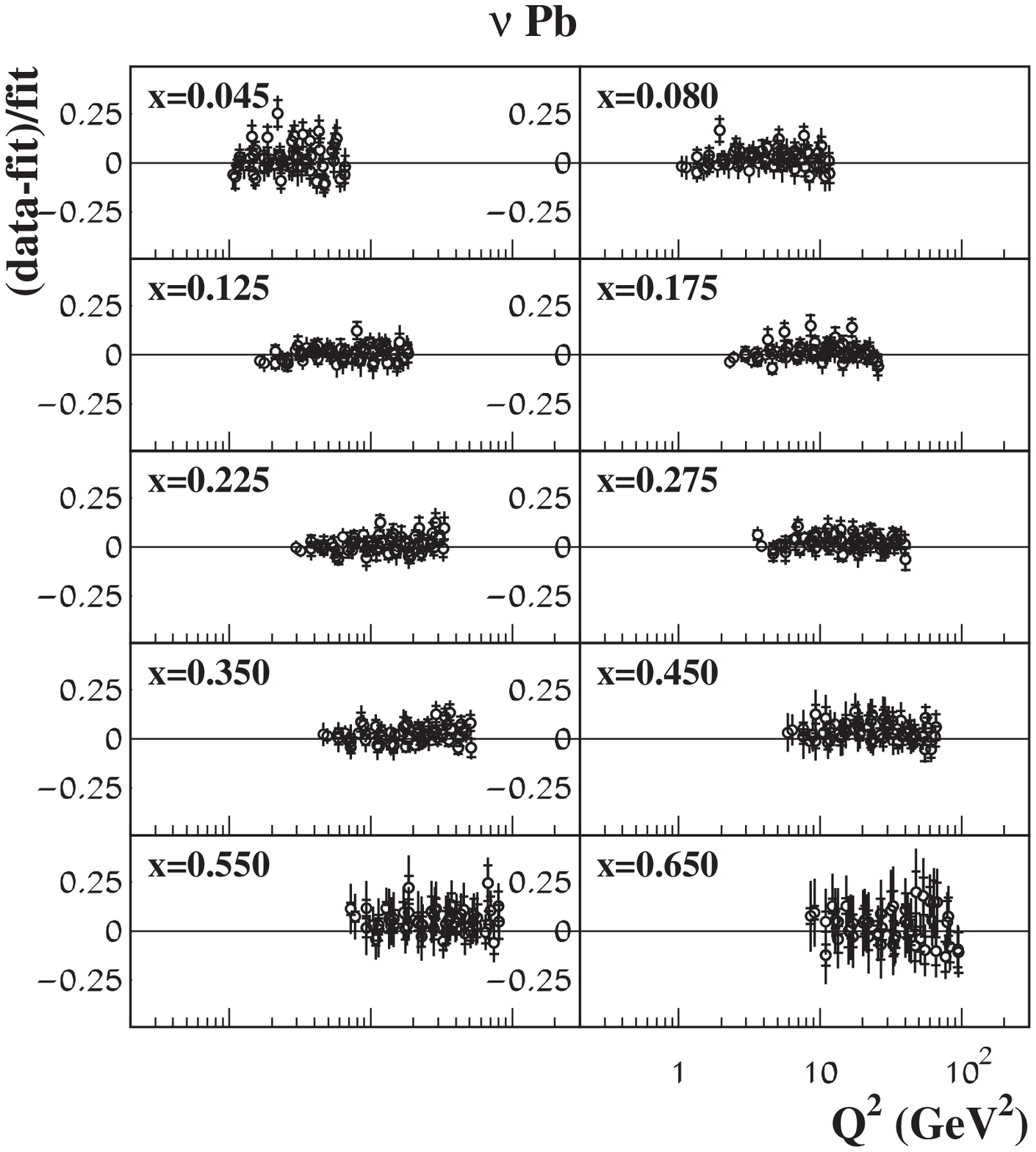}\includegraphics[height=.40\textheight]{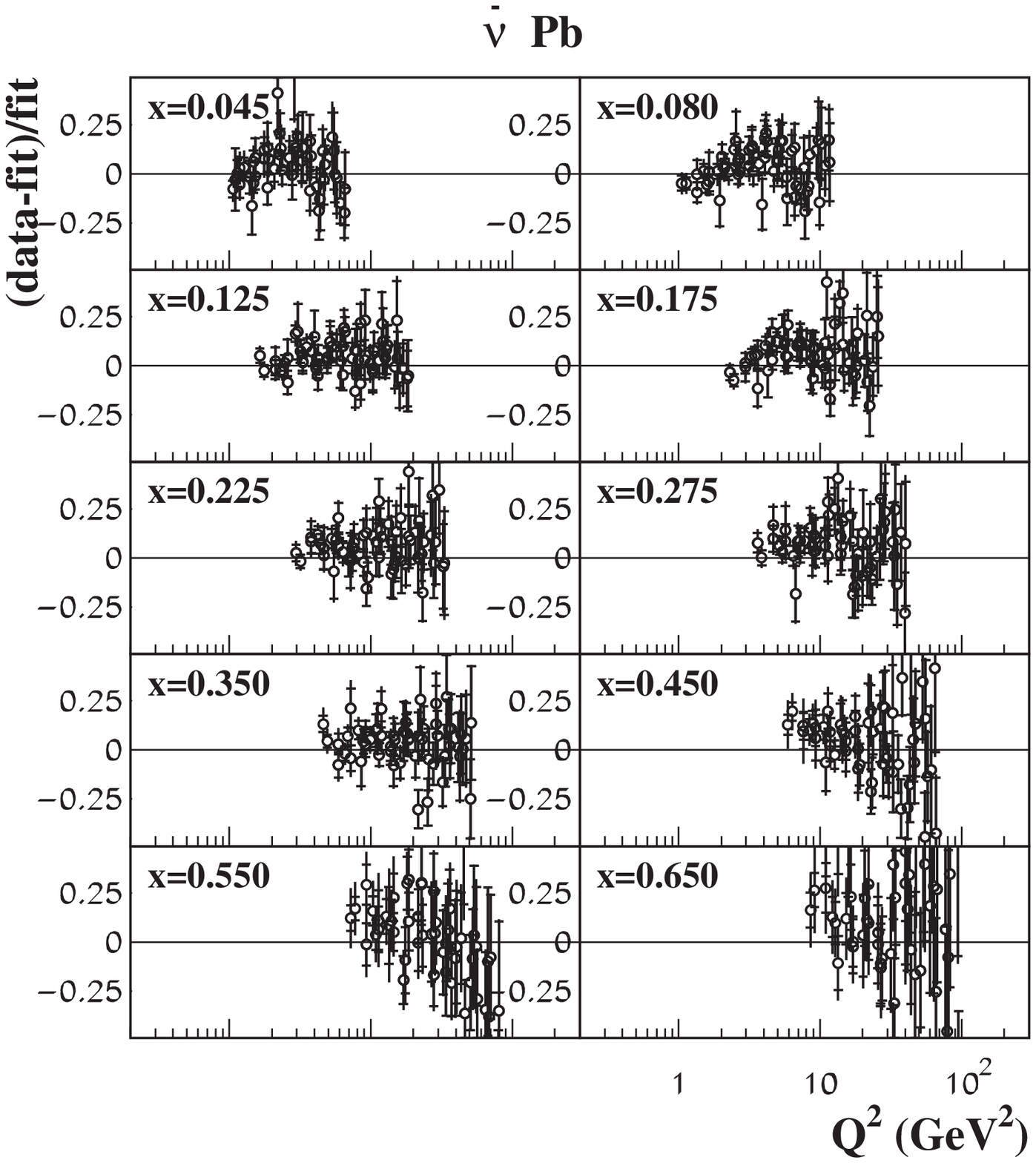}
  \caption{Pulls for neutrino (left) and antineutrino (right) cross-section data from CHORUS~\cite{chorus-xsec} 
with respect to our calculations as a function of $Q^2$. Nuclear corrections~\cite{Kulagin:2004ie} for the 
lead target and electroweak corrections~\cite{Arbuzov-Bardin} are taken into account.   
}\label{fig:nupulls}  
\end{figure}

Figure~\ref{fig:nupulls} shows the pulls of our fit to neutrino and 
antineutrino inclusive inelastic cross-section data from the CHORUS experiment~\cite{chorus-xsec} 
on a lead target. The calculation includes nuclear corrections~\cite{Kulagin:2004ie} for the lead 
target, electroweak corrections~\cite{Arbuzov-Bardin} and the PCAC contribution.    

The determination of LT and HT terms is performed from all available
data with $Q^2>1.0~GeV^2$ and $W>1.8~GeV$. It is interesting to check the
extrapolation of DIS structure functions into the resonance region for $W<1.8$ GeV.
Our results are consistent with the duality principle and the  
integral of the difference between the recent JLab resonance data 
and the average DIS predictions is consistent with zero within few 
percent in the entire kinematic region. This can be also considered as an indirect indication  
in favour of a negligible twist-6 contribution to structure functions.


\begin{theacknowledgments}

This work is partially supported by the RFBR grant 06-02-16659 and the Russian
Ministry of Science and Education Nsh 5911.2006. R.P. thanks USC for supporting this research.

\end{theacknowledgments}


\begin{thebibliography}{9}

\bibitem{NNLO}
  S.~Moch, J.~A.~M.~Vermaseren and A.~Vogt,
  \emph{Nucl.\ Phys.\  B} \textbf{688}, 101 (2004); 
  A.~Vogt, S.~Moch and J.~A.~M.~Vermaseren,
  \emph{Nucl.\ Phys.\  B} \textbf{691}, 129 (2004);
  J.~A.~M.~Vermaseren, A.~Vogt and S.~Moch,
  \emph{Nucl.\ Phys.\  B} \textbf{724}, 3 (2005);
  S.~Moch and M.~Rogal,
  arXiv:0704.1740 [hep-ph].

\bibitem{Miramontes:1989ni}
  J.~L.~Miramontes, M.~A.~Miramontes and J.~Sanchez Guillen,
  \emph{Phys.\ Rev.\  D} \textbf{40}, 2184 (1989).

\bibitem{Rslac}  
  L.~W.~Whitlow et al.,  
  \emph{Phys.\ Lett.\  B} \textbf{250}, 193 (1990).  

\bibitem{Virchaux92}
  M.~Virchaux and A.~Milsztajn,
  \emph{Phys.\ Lett.\  B} \textbf{274}, 221 (1992).

\bibitem{BodekYang00}
  U.~K.~Yang and A.~Bodek,
  \emph{Eur.\ Phys.\ J.\ C} \textbf{13}, 241 (2000).

\bibitem{Alekhin02}
  S.~I.~Alekhin,
  \emph{Phys.\ Rev.\ D} \textbf{68}, 014002 (2003).

\bibitem{Kataev99}
  A.~L.~Kataev, G.~Parente and A.~V.~Sidorov,
  \emph{Nucl.\ Phys.\ B} \textbf{573}, 405 (2000).

\bibitem{TMC}
  O.~Nachtmann,
  \emph{Nucl.\ Phys.\  B} \textbf{63}, 237 (1973);
  H.~Georgi and H.~D.~Politzer,
  \emph{Phys.\ Rev.\  D} \textbf{14}, 1829 (1976).

\bibitem{Alekhin:2006zm}  
  S.~Alekhin, K.~Melnikov and F.~Petriello,  
  \emph{Phys.\ Rev.\  D} \textbf{74}, 054033 (2006).  

\bibitem{Adams:1996gu}
  M.~R.~Adams et al.  [E665 Collaboration],
  \emph{Phys.\ Rev.\  D} \textbf{54}, 3006 (1996).

\bibitem{Kulagin:2004ie}
  S.~A.~Kulagin and R.~Petti,
  \emph{Nucl.\ Phys.\  A} \textbf{765}, 126 (2006).  

\bibitem{chorus-xsec}
  G. \"{O}neng\"{u}t et al. [CHORUS Collaboration],
  \emph{Phys. Lett. B} \textbf{632}, 65 (2006).

\bibitem{E118}
   C.~Keppel, private communication.

\bibitem{BraunKolesnichenko87}  
  V.~M.~Braun and A.~V.~Kolesnichenko,
  \emph{Nucl.\ Phys.\  B} \textbf{283}, 723 (1987).

\bibitem{Blumlein:2006be}
  J.~Bl\"{u}mlein, H.~Bottcher and A.~Guffanti,
  \emph{Nucl.\ Phys.\  B} \textbf{774}, 182 (2007).

\bibitem{Zeller:2001hh}
  G.~P.~Zeller et al. [NuTeV Collaboration],
  \emph{Phys.\ Rev.\ Lett. } \textbf{88}, 091802 (2002)
  [Erratum-ibid.\  {\bf 90}, 239902 (2003)].

\bibitem{Kulagin:2003wz}
  S.~A.~Kulagin,
  \emph{Phys.\ Rev.\  D} \textbf{67}, 091301 (2003).

\bibitem{Kulagin:2007ie}
  S.~A.~Kulagin and R.~Petti,
  arXiv:hep-ph/0703033.  

\bibitem{Yang:2001xc}
  U.~K.~Yang {\it et al.}  [CCFR/NuTeV Collaboration],
  \emph{Phys.\ Rev.\ Lett.}  \textbf{87}, 251802 (2001).

\bibitem{ccfr-pcac}
  B.~T.~Fleming et al. [CCFR Collaboration],
  \emph{Phys.\ Rev.\ Lett.}  \textbf{86}, 5430 (2001).

\bibitem{Arbuzov-Bardin}
  A.~B.~Arbuzov, D.~Yu.~Bardin and L.~V.~Kalinovskaya,  
  \emph{JHEP} 0506:078 (2005).  


\end{thebibliography}
\end{document}